\documentclass[11pt,twoside]{article}
\usepackage[pdftex]{graphicx}
\usepackage{amsmath}
\usepackage{amssymb}
\usepackage{cite}

\usepackage{bm,bbold}
\usepackage{hyperref}

\begin{document}
\newcommand{\pst}{\hspace*{1.5em}}

\newcommand{\rigmark}{\em Journal of Russian Laser Research}
\newcommand{\lemark}{\em Volume 30, Number 5, 2009}

\newcommand{\ds}{\displaystyle}
\newcommand{\bea}{\begin{eqnarray}}
\newcommand{\eea}{\end{eqnarray}}
\newcommand{\arcsinh}{\mathop{\rm arcsinh}\nolimits}
\newcommand{\arctanh}{\mathop{\rm arctanh}\nolimits}
\newcommand{\bc}{\begin{center}}
\newcommand{\ec}{\end{center}}


\def\ba{\begin{equation}\begin{array}{c}}
\def\ea{\end{array}\end{equation}}
\def\be{\ba\displaystyle}
\def\ee{\ea}

\newcommand{\ra}{\rangle}
\newcommand{\la}{\langle}

\newcommand{\sign}{{\rm sign}}

\newcommand{\GS}{{\rm GS}}
\renewcommand{\H}{\hat H}
\newcommand{\V}{\hat V}
\newcommand{\kF}{k_{\rm F}}
\newcommand{\vF}{v_{\rm F}}
\newcommand{\kB}{k_{\rm B}}
\newcommand{\vc}{v_{\rm c}}
\newcommand{\vs}{v_{\rm s}}
\newcommand{\uR}{u_{\rm R}}
\newcommand{\uL}{u_{\rm L}}
\newcommand{\wa}{w_{\rm a}}

\newcommand{\q}{\mathbf q}

\thispagestyle{plain}

\label{sh}


\begin{center} {\Large \bf
\begin{tabular}{c}
A necessary condition for quantum adiabaticity 
\\[-1mm]
applied to the adiabatic Grover search
\end{tabular}
 } \end{center}

\bigskip

\bigskip

\begin{center} {\bf
Oleg Lychkovskiy$^{1,2,\,*}$
}\end{center}

\medskip

\begin{center}
{\it
$^1$Skolkovo Institute of Science and Technology,\\
Skolkovo Innovation Center 3, Moscow  143026, Russia

\smallskip

$^2$Steklov Mathematical Institute of Russian Academy of Sciences,\\
Gubkina str. 8, Moscow 119991, Russia
}
\smallskip

$^*$Corresponding author e-mail:~~~o.lychkovskiy@skoltech.ru\\
\end{center}

\begin{abstract}\noindent
Numerous sufficient conditions for adiabaticity of the evolution of a driven quantum system have been known for quite a long time. In contrast, necessary adiabatic conditions are scarce. A practicable necessary condition well-suited for many-body systems has been proven recently  in \href{https://doi.org/10.1103/PhysRevLett.119.200401}{[Phys. Rev. Lett. 119, 200401 (2017)]}. Here we tailor this condition for estimating run times of adiabatic quantum algorithms. As an illustration, the condition is applied to the adiabatic algorithm for searching in an unstructured database (adiabatic Grover search algorithm). We find that thus obtained lower bound on the run time of this algorithm reproduces $\sqrt N$ scaling ($N$ being the number of database entries) of the explicitly known optimal run time. This observation highlights the merits of the new adiabatic condition and its potential relevance to adiabatic quantum computing.
\end{abstract}

\medskip

\noindent{\bf Keywords:}
quantum adiabatic theorem, adiabatic conditions, adiabatic quantum algorithms, adiabatic Grover search.

\section{Preliminaries and motivation}
\pst
Consider a quantum system with a Hamiltonian $\H_\lambda$, where $\lambda=\lambda(t)$ is a time-dependent parameter.  For each $\lambda$ one can define an instantaneous ground state, $\Phi_\lambda$,
which is the lowest eigenvalue solution to the Schr\"odinger's stationary equation,
\begin{equation}\label{eigenproblem}
\H_\lambda\, \Phi_\lambda = E_\lambda \, \Phi_\lambda.
\end{equation}
Here $E_\lambda$ is the instantaneous ground state energy. We assume that the ground state is non-degenerate for any~$\lambda$.
The dynamics of the system is governed by the Schr\"odinger  equation, which can be written in a rescaled form,
\begin{equation}\label{Schrodinger equation}
i\,\dot{\lambda}\, \partial_\lambda \Psi_\lambda = \H_{\lambda}\,\Psi_\lambda.
\end{equation}
Here $\dot{\lambda}\equiv\partial_t \lambda(t)$ and $\Psi_\lambda$ is the state vector of the system which depends on time through the time-dependent parameter $\lambda$.  Initially the system is prepared in the instantaneous ground state:
\begin{equation}\label{initial condition}
 \Psi_0= \Phi_0.
\end{equation}

The functional dependence of $\lambda$ over $t$ is called a {\it schedule}. We define a schedule on some finite time interval, $t\in [0,t_f]$. Without loss of generality, we take
\be\label{lambda}
\lambda(0)=0 ~~~~{\rm  and}~~~~ \lambda(t_f)=1.
\ee
Time $t_f>0$ is called the run time of the schedule. We assume that $\lambda(t)$ is smooth. We further assume that $\lambda$ is non-decreasing, which implies that
\be\label{lambda max}
\lambda_t \leq 1, ~~~ t\in [0,t_f].
\ee

Any schedule can be rescaled, i.e. a new schedule $\tilde\lambda(t)= \lambda(\eta t)$ can be defined, where $\eta>0$ is the scaling factor and $\tilde t_f=t_f/\eta$ is the  run time of the rescaled schedule. We will refer to all possible rescalings of a given schedule as a {\it path}. 
In other words, a path is a family of schedules which can be transformed one into another by rescaling the time. Notice that we allow only linear rescalings in this definition. Obviously, for a given path,  picking  a specific schedule $\lambda(t)$  amounts to fixing the run time, $t_f$, and vice versa.


The evolution is called adiabatic as long as the state of the system, $\Psi_\lambda$, stays close to the instantaneous ground state,  $\Phi_\lambda$. The celebrated {\it quantum adiabatic theorem}  \cite{born1926,born1928beweis,kato1950} states that for  however small {\it allowance} $\epsilon\in[0,1]$ and arbitrary path
there exists a run time $t_f$ large enough that
\be\label{adiabatic allowance}
1-\mathcal F(1)<\epsilon,
\ee
where the adiabatic fidelity,
\begin{equation}\label{fidelity}
 \mathcal F(\lambda) \equiv \left | \langle \Phi_\lambda
 \vert \Psi_\lambda \rangle\right|^2,
\end{equation}
quantifies how close $\Psi_\lambda$ and $\Phi_\lambda$ are.

Much of the practical value of the adiabatic theorem depends on the availability of estimates on the adiabatic run time, $t_f$, which do not require explicitly solving the time-dependent Schr\"odinger equation. In fact, the adiabatic theorem is proved by deriving a {\it sufficient} condition on the run time, $t_f$, which ensures that the
eq. \eqref{adiabatic allowance} is satisfied for a given allowance $\epsilon$. Numerous sufficient adiabatic conditions have been proved since the quantum adiabatic theorem had been introduced, see e.g. ref. \cite{albash2016adiabatic} for a review.

Clearly, it is highly desirable to complement  sufficient adiabatic conditions by {\it necessary}  adiabatic conditions, i.e. lower bounds on $t_f$ provided inequality \eqref{adiabatic allowance}  holds. Quite remarkably, practical  necessary adiabatic condition are scarce. In fact, to the best of our knowledge, the only known way to get such conditions is to use some form of a quantum speed limit, which is an advanced rigorous version of the time-energy uncertainty relation \cite{pfeifer1995generalized,salem2007adiabatic,deffner2017quantum}.\footnote{A number of necessary adiabatic conditions of different nature  have also been reported \cite{tong2010quantitative,rigolin2012adiabatic,boixo2010necessary,wang2016necessary}, however, we believe that they lack either convincing justification or practicability. A condition of ref. \cite{tong2010quantitative} (see also its generalization in ref. \cite{rigolin2012adiabatic}) has triggered a debate \cite{zhao2011comment,comparat2011comment,tong2011tong}, and we agree with ref. \cite{zhao2011comment} that unjustified omission of certain terms undermines the result of  \cite{tong2010quantitative}. Difficulties of the approach of \cite{tong2010quantitative} are clearly exposed in ref. \cite{li2014why}, where a rigorous bound of the type \cite{tong2010quantitative} is obtained, which, however, lacks practicability since it requires the knowledge of $\Psi_\lambda$. A lower bound on  $(t_f \Delta)$, where $\Delta$ is the minimal energy gap between the ground state and the rest of the spectrum along the path, has been reported in ref. \cite{boixo2010necessary}. We remark, however, that one can easily construct a simple example where this bound is violated but the quantum evolution is adiabatic \cite{somma2012quantum}. To this end one can introduce an auxiliary eigenstate which is dynamically decoupled from the ground state (e.g. due to some symmetry) but has energy separated from the ground state energy by an arbitrarily small gap (more sophisticate counterexamples can also be constructed \cite{somma2012quantum}). Thus the result of ref. \cite{boixo2010necessary} does not constitute a valid necessary adiabatic condition. The result of ref. \cite{wang2016necessary} does not provide a lower bound on $t_f$ for a given path and a given allowance $\epsilon$ and thus, in our view, is of limited utility.} A necessary condition of this type well suited for many-body systems has been proven recently in ref. \cite{lychkovskiy2017time}. This condition has been successfully applied to two adiabatically driven systems. The first one is the topological Thouless pump \cite{thouless1983quantization}, where a necessary condition for the transport quantization in a continuous mode of operation has been established \cite{lychkovskiy2017time}. The second one is a mobile impurity driven through a one-dimensional fluid, where a necessary condition for adiabatic quasi-Bloch oscillations \cite{gangardt2009bloch} has been derived \cite{lychkovskiy2018quantum}, thus resolving a long-standing controversy \cite{gamayun2014kinetic,schecter2015comment,gamayun2015reply}.

Above-mentioned previous applications of the necessary adiabatic condition \cite{lychkovskiy2017time,lychkovskiy2018quantum} concern problems in condensed matter physics. The purpose of the present paper is to tailor this condition for estimating the run time of an adiabatic quantum algorithm and to illustrate how it works for a paradigmatic quantum algorithm --- the adiabatic Grover search \cite{roland2002quantum}.



The rest of the paper is organized as follows. In the next section we briefly review the central result of ref. \cite{lychkovskiy2017time} and shape it in the form of the lower bound on the  run time of a quantum adiabatic algorithm. Then we apply this bound to the adiabatic Grover search algorithm.
We conclude by discussing how the established bound compares to the exact result for the Grover search.




\section{Necessary adiabatic condition}
\pst
The central result of Ref. \cite{lychkovskiy2017time} is an inequality which binds the adiabatic fidelity,  $\mathcal F_\lambda$, to the overlap between the instantaneous eigenstates,
\be
\mathcal C_\lambda \equiv
\vert \langle  \Phi_{\lambda}
 \vert \Phi_{0}\rangle \vert^2.
\ee
Here we quote this inequality for a special form of the Hamiltonian important for the adiabatic quantum computing,
\be\label{H}
\H_\lambda=\H_0+\lambda \, (\H_1-\H_0).
\ee
For this Hamiltonian  the inequality reads \cite{lychkovskiy2017time}
\begin{equation}
\label{inequality}
|{\cal F}_\lambda-{\cal C}_\lambda|  \leq
\delta V_N\,\int_0^{t_f} \lambda(t)\,dt\, ,
\end{equation}
where $\delta V_N$ is the uncertainty of the driving term, $\V\equiv\H_1-\H_0$, with respect to the initial state of the system,~$\Phi_0$,
\be
\delta V_N \equiv \sqrt{\langle \hat V^2 \rangle_0
- \langle   \hat V \rangle_0^2},
\ee
with $\langle\dots\rangle \equiv \langle \Phi_0 |\dots| \Phi_0 \rangle$. The subscript $N$ in $\delta V_N$ emphasises that $\delta V_N$ in general depends on the system size,~$N$.

The inequality \eqref{inequality} along with  eqs. \eqref{lambda max} and \eqref{adiabatic allowance} leads to the following necessary condition for adiabaticity with the allowance $\epsilon$:
\be\label{necessary condition}
t_f\geq \frac{1-\epsilon-{\cal C}(1)}{\delta V_N }.
\ee
This is the first main result of the paper.

It should be noted that the above bound is somewhat different from the necessary adiabatic condition established in refs. \cite{lychkovskiy2017time,lychkovskiy2018quantum}. The reason is that the properties of ${\cal C}_\lambda$ has been exploited in addition to the inequality~\eqref{inequality} in \cite{lychkovskiy2017time,lychkovskiy2018quantum}. This has become possible since ${\cal C}_\lambda$ quantifies the generalized orthogonality catastrophe which is well understood in the many-body context. However, no such additional information is in general available in the adiabatic quantum computing context.

\section{Adiabatic Grover search}
\pst
The idea of the adiabatic quantum computing is to translate the input of a certain problem into couplings of a physically realizable Hamiltonian, $\H_1$, in such a way that the ground state of this Hamiltonian encodes the solution to the problem, and then to prepare this unknown ground state starting from another, known and easily preparable ground state of another Hamiltonian, $\H_0$, by adiabatically interpolating between the two Hamiltonians \cite{farhi2000quantum,farhi2001quantum,albash2016adiabatic}. The later step can be realized with a time-dependent Hamiltonian of the form \eqref{H}.\footnote{For the description of the former step we refer the interested reader to the review \cite{albash2016adiabatic} and references therein.}
The specific forms of $H_0$, $H_1$ and $\lambda(t)$ are reffered to as a {\it quantum adiabatic algorithm}.
In this context $t_f$ stands for the run time of the algorithm, ${\cal F}_1$ -- for the success probability of the computation and $\epsilon$ - for the upper bound on the probability of an error.

Here we consider a particularly well-studied quantum adiabatic algorithm  aimed at the search of a marked element in an unstructured database with $N$ entries (adiabatic Grover search)  \cite{roland2002quantum}.  It can be realized with the Hamiltonians
\be\label{H_P and H_B}
\H_1=\mathbb{1}-|m \rangle\langle m|~~~{\rm and}~~~\H_0=\mathbb{1}-|\chi\rangle\langle \chi|,
\ee
acting in $N$-dimensional Hilbert space,  where $\mathbb{1}$ is the unit operator, $\{|j\rangle\},\, j=1,...N$ is an orthonormal basis, $m$ is the number of the marked database entry to be searched for, and
\be
|\chi\rangle \equiv \frac 1 {\sqrt N}\sum_{j=1}^N|j\rangle.
\ee

We refer the reader to the original paper \cite{roland2002quantum} and the review \cite{albash2016adiabatic} for the quantum information context of the problem, and focus here on conditions for adiabatically preparing the ground state of $\H_1$. In fact, the central question in the field of adiabatic quantum computing is  how the run time of an adiabatic algorithm, $t_f$, scales with the size of the input, $N$, in the best case (i.e. for an optimal schedule $\lambda(t)$ which provides the minimal $t_f$ for a given error $\epsilon$).
For many quantum adiabatic algorithms this question is a matter of debate \cite{albash2016adiabatic}. For the adiabatic Grover algorithm, however, the answer is known and reads \cite{roland2002quantum,albash2016adiabatic}
\be\label{scaling}
t_{f\,\,min}=O\left(\sqrt N\right).
\ee
Here we demonstrate that the lower bound \eqref{necessary condition} reproduces this scaling.

Indeed, a straightforward calculations gives
\be
\delta V_N=\frac1{\sqrt N} \sqrt{1-\frac1N},
\ee
\be
{\cal C}(1) =\frac1N.
\ee
Substituting these results into eq. \eqref{necessary condition} one obtains the lower bound for the run time of the adiabatic Grover search performing with the error below $\epsilon$:
\be\label{main result}
t_f\geq  \sqrt{N}\, (1-\epsilon)\,  \frac{1-N^{-1}(1-\epsilon)^{-1}}{\sqrt{1-N^{-1}}} .
\ee
This bound is the second main result of the paper.
Clearly, it scales as $\sqrt{N}$ in the limit of large $N$.

\section{Discussion}
\pst
The fact that the scaling of the lower bound~\eqref{main result}  coincides with the optimal scaling is quite remarkable.  It should be contrasted to the poor performance of the sufficient adiabatic conditions which guaranty adiabaticity only for $t_f=O(N)$ (this does not constitute any speedup over the classical database search) \cite{roland2002quantum,albash2016adiabatic}. While the exact optimal scaling has been obtained in ref. \cite{roland2002quantum} by thoroughly examining the spectrum  of the parameter-dependent Hamiltonian \eqref{H}, the same scaling is recovered here from the necessary adiabatic condition with much less technical effort and without any knowledge of the spectrum (and, in particular, the minimal gap). The latter fact is of particular importance since the spectrum of $\H_\lambda$ is not known for a typical quantum algorithm. We conclude that the necessary adiabatic condition \eqref{necessary condition}  has a potential to become  an efficient tool in evaluating the performance of quantum adiabatic algorithms.

\medskip
\noindent{\it Note added.}~ After this paper had been finalized I became aware of a similar results obtained in refs. \cite{kieu2017new,kieu2018travelling}. In ref. \cite{kieu2017new} a necessary adiabatic condition similar to the condition \eqref{necessary condition} is reported.
In ref. \cite{kieu2018travelling} the condition of ref. \cite{kieu2017new} is applied to the adiabatic search algorithm.

\section*{Acknowledgments}
\pst
I am grateful to V. Cheianov and O. Gamayun for fruitful discussions. The work was supported by the Russian Science Foundation under the grant N$^{\rm o}$ 17-71-20158.

\end{document}